# Explosion cratering in 3D granular media

Tianyu Liu, Boen Cao, Xiao Liu, Ting-Pi Sun, and Xiang Cheng*

*Department of Chemical Engineering and Materials Science, University of Minnesota, Minneapolis, MN 55455, USA*

*Email: xcheng@umn.edu

Sudden release of energy in explosion creates craters in granular media. In comparison with well-studied impact cratering in granular media, our understanding of explosion cratering is still primitive. Here, we study low-energy lab-scale explosion cratering in 3D granular media using controlled pulses of pressurized air. We identify four regimes of explosion cratering at different burial depths, which are associated with distinct explosion dynamics and result in different crater morphologies. We propose a general relation between the dynamics of granular flows and the surface structures of resulting craters. Moreover, we measure the diameter of explosion craters as a function of explosion pressures, durations and burial depths. We find that the size of craters is non-monotonic with increasing burial depths, reaching a maximum at an intermediate burial depth. In addition, the crater diameter shows a weak dependence on explosion pressures and durations at small burial depths. We construct a simple model to explain this finding. Finally, we explore the scaling relations of the size of explosion craters. Despite the huge difference in energy scales, we find that the diameter of explosion craters in our experiments follows the same cube root energy scaling as explosion cratering at high energies. We also discuss the dependence of rescaled crater sizes on the inertial numbers of granular flows. These results shed light onto the rich dynamics of 3D explosion cratering and provide new insights into the general physical principles governing granular cratering processes.

# 1 Introduction

Craters in granular media are ubiquitous in nature. Raindrop dimples on beach,[1-4] potholes in sedimentary bed[5,6] and volcanic craters and asteroid impact craters in various planetary bodies[7,8] are a few examples of granular craters across widely different length and energy scales. Understanding the process of granular cratering is important not only for the fundamental geological research but also for various technical applications such as spraying coating of powders and grains,[9] the design of explosives[10] and the manufacture of shock and energy dampening devices in automotive and aerospace industries.[11] Explosion and impact are by far the two most common ways to create craters in granular media. Indeed, when Robert Hooke discussed the origin of craters on the Moon in his famous book *Micrographia*,[12] he first proposed impact as the possible origin, then rejected his own hypothesis and suggested the explosion of rising subterranean bubbles as the origin of lunar craters. In spite of the mistake, we still acknowledge Hooke as the first person who elucidated the true origin of planetary craters today.[7]

The relation between explosion cratering and impact cratering is profound and has been explored at high-energy scales in planetary science.[7,13] At high energies associated with hypervelocity impacts of asteroids, the so-called impact-explosion analogy has been proposed, where a hypervelocity impact moving more than a few kilometers per second is quantitatively likened to an explosion of comparable energy at a certain empirically derived burial depth.[7,14] The analogy greatly facilitates the study of asteroid impact cratering using high-energy explosions such as those from nuclear weapons tests.[15,16] At low energy scales, although impact cratering has been extensively investigated in granular physics,[17-19] the study of low-energy explosion cratering is still few and far between. Most of the existing lab-scale experiments focused on two-dimensional (2D) systems and/or near-surface explosions for the convenience of imaging.[20,21] Pacheco-Vazquez and co-workers investigated near-surface explosions using firecrackers, where they compared the morphology of explosion craters with that of impact craters.[20] In particular, they found that the diameter of explosion craters follows a power-law scaling with explosion energy as $E^{0.3}$. The Pacheco-Vazquez group also studied three-dimensional (3D) granular cratering induced by the collapse of underground cavities.[22] They showed that the crater size depends solely on the volume of cavities and furthermore investigated the dynamics of the resulting granular jets. Energy scaling was not reported in this study. Our own previous work focused on explosion cratering in quasi-2D granular bed.[21] We identified different regimes of explosion cratering, where different energy scaling rules were discovered. A 1/3 scaling similar to that in Ref. 20 is found in the so-called bubbling regime. More recently, Marston and Pacheco-Vazquez studied granular cratering induced by pulsed laser ablation on the surface of granular bed.[23] Millimetric craters formed in the ablation process shows a range of energy-scaling exponents from 0.31 up to 0.43, depending on the properties of granular particles.

Although these pioneering studies have started to reveal the rich dynamics of low-energy explosion cratering, many important aspects of cratering processes remain unexplored. Particularly, explosion cratering in 3D granular media at varying burial depths—a key control parameter of explosion—has not been studied. In this study, we investigate explosion cratering in 3D granular media induced by well-controlled pulses of pressurized air. We systematically explore explosion dynamics and the morphology of explosion craters as a function of explosion pressures, durations and burial depths. Four different explosion regimes are identified in our experiments, which exhibit different cratering dynamics and crater morphologies. Based on the observation across different regimes, we generalize a relation between the dynamics of granular flows and the surface morphologies of resulting craters, applicable to both explosion and impact granular cratering. We further analyze the diameters of explosion craters, which show a non-monotonic dependence on burial depths. A maximal crater occurs at an optimal intermediate burial depth. At shallow burial depths, we reveal a geometric origin of the size of craters. We also investigate the scaling relations of crater morphology, in particular the energy scaling of crater sizes, via dimensional analyses. We find that the diameter of granular craters follows the 1/3 scaling with explosion energy, similar to that discovered in high-energy explosions, but different from the familiar 1/4 energy scaling of low-energy impact cratering. These results provide a quantitative understanding of explosion cratering in 3D systems and extend our knowledge on the energy scaling of general granular cratering processes. Our study also supports the derivation of the explosion energy of pulsed pressurized gases in granular media and serves as the first step to seek the relation between impact cratering and explosion cratering in the low-energy regime.

## 2 Experiments

The major component of our experimental setup is a transparent cylindrical polyethylene terephthalate (PET) container of an inner diameter of 13.9 cm and a height of 29.7 cm (Fig. 1). The inner wall of the container is coated with anti-static spray to avoid the accumulation of electrostatic charges on granular particles. We design an automatic air jetting system, which can deliver short pulses of high-pressure air over controlled time intervals. The opening of the jetting system is a stainless steel nozzle with an inner diameter of 0.56 mm. Supported by a T-shaped stand, the nozzle points vertically upward against gravity and is centered 20.8 cm below the top of the container. A fine Nylon mesh of a nominal sieve opening of 75 µm is glued to the opening to prevent the falling of granular particles into the nozzle. The nozzle is further connected to a compressed-air tank via a fast-response solenoid valve (Parker Hannifin Series 99). The valve opening time, $\tau$, can be finely controlled by a function generator via a custom Matlab program. Since the volume of the tank is much larger than the total volume of air pulses used in our experiments, the explosion pressure, $P$, can be treated as a constant in a given explosion event, which is set by the pressure in the tank. We vary $P$ between $2.1 \times 10^6$ Pa and $8.3 \times 10^6$ Pa in our experiments and fix $\tau = 20$ ms unless stated otherwise. Soda-lime glass beads of mean diameter of 90 µm are used as our granular particles (Mo-Sci Corporation, Class IV Microspheres). The polydispersity of the particles is under 17% from direct measurements. The granular particles are filled to a controlled height $d$ above the opening of the nozzle, which gives the burial depth of explosions. We vary $d$ between 0 and 125 mm in our experiments. An adjustable leveler is used to flatten the surface of granular bed before each experiments. Finally, a high-speed camera (Photron SA-X2) is positioned in front of the container to image the dynamics of explosion at the frame rate of 1000 frames per second. The camera is synchronized with the open of the valve through an electrical trigger signal.

In a typical experiment, an explosion is triggered in a granular bed. Triggered by the same signal, the high-speed camera takes the images of the cratering process simultaneously (Fig. 2). After explosion, an approximately circular crater forms on the surface of the granular bed (Fig. 3). To determine the size of the resulting crater, we illuminate the crater with a shallow light beam about 10° above the flat granular surface. Such a small angle highlights the rising rim of the crater. A high-resolution static image is then taken by a separate camera. The average diameter of the crater is obtained by fitting the crater with a circle using the Fiji software. For craters with clear rims, the fitting circles align with the peak of the rims. For craters without rims, crater diameters are determined by the locations where deformations from the flat surface can be observed. Naturally, the diameter of craters without rims is subject to relatively larger measurement errors. After each measurement, we remove the top layer of granular bed and thoroughly stir the entire bed manually. Finally, after adjusting the amount of granular particles and the height of the leveler to the intended burial depth, we sweep the leveler across the bed multiple times, so that the initial surface is flat above the explosion site and normal to the direction of gravity.

## 3 Results and discussion

### 3.1 Explosion dynamics

The dynamics of explosion cratering in 3D granular media are richer and more complicated than that of 2D explosion cratering.[21] We identify four different explosion-cratering regimes at different burial depths $d$ in our experiments, which we shall discuss below in the sequence of increasing $d$ with a fixed explosion pressure $P = 4.1 \times 10^6$ Pa and an explosion duration $\tau = 20$ ms.

(*i*) Shallow air-blow regime. When $d$ is below 5 mm, the granular layer above the explosion site is shallow. Upon the open of the valve, particles above the nozzle are immediately carried away by the strong air flow escaping the nozzle and scatter randomly over a large area of the surface of granular bed (Fig. 2a). Moving ballistically, the particles do not form a clear granular corolla, typically observed in explosion cratering at deeper burial depths and in impact cratering.[24] The resulting craters are small with a smooth conical wall without rims (Fig. 3a). The dynamics of explosion in this regime are dominated by the kinetic energy of the air pulse as we shall discuss below in Sec. 3.3.

(*ii*) Eruption regime. When $d$ is above 10 mm and below ~ 55 mm, eruption cratering similar to that of 2D explosion cratering is observed (Fig. 2b). In this regime, particles expelled by the explosion move collectively and form a clear granular corolla with a well-defined boundary. The wall of the corolla forms an acute angle with the surface of granular bed at small $d$ (Fig. 2b), which approaches 90° towards the end of the eruption regime at larger $d$. Thanks to the expansion of the corolla, which pushes particles sideways away from the explosion site along the surface of granular bed, pronounced rim can be observed around explosion craters (Fig. 3b). The falling of corolla partially fills the outer region of the crater near the rim. The crater in this regime has a shallow depth and does not possess a smooth conical shape. When $d$ is above 25 mm, a granular jet can be observed in the middle of the falling granular corolla (Fig. 2c). The jet forms due to the collapse of the cavity created by the explosion.[22,25-29] Without the confining walls, the jet is much stronger than that formed in 2D explosion.[21] The falling of the jet gives rise to an irregular hump at the center of the crater (Fig. 3c). When $d$ is above 40 mm, the time it takes for the surface first shows deformation becomes comparable to the explosion duration (Fig. 4). Eruption gradually transitions to bubbling as we shall discuss next. In the eruption regime, the expansion of pressurized air quantified by the internal energy of the air pulse assumes the leading role in controlling the dynamics of explosion (Sec. 3.3).

(*iii*) Bubbling regime. As expected from the study of 2D explosion,[21] at even larger $d$, bubbling cratering replaces eruption cratering. Nevertheless, the transition between the two regimes is gradual. For $d$ between 40 mm and 55 mm, cratering dynamics show a mixed feature of eruption and bubbling (Fig. 2c). A granular dome first forms above the granular bed similar to the bubbling explosion as discussed below. Then an eruption occurs on the top of the dome and forms a granular corolla with an almost vertical wall. The falling of the vertical granular corolla creates thick rims around the crater, whereas the falling of the granular jet gives rise to the central hump in the crater (Fig. 3c). Above 55 mm, a clear bubbling behavior can be identified (Fig. 2d): because of the deep burial depth, the valve closes before the bubble containing high-pressure air reaches the surface of granular bed. The underground bubble then rises to the surface and pushes the granular bed into a dome-like structure. The bubble, as well as the dome, finally collapses without eruption and leaves an explosion crater in the granular bed. Although without eruption a granular corolla cannot be observed, a thin granular jet still forms due to the collapse of the bubble. A weak but sharp rim is formed around the crater from the collapse of the bubble (Fig. 3d). The falling of the thin jet results in a small hump in the center of the crater. As $d$ increases, crater size decreases in the bubbling regime.

(*iv*) Sink-in regime. At large $d$ above 65 mm, the surface deformation of granular bed is significantly delayed compared to explosion duration (Fig. 4). The surface of the bed bulges upward slightly and then sinks back, leaving a caved-in crater (Fig. 2e). In contrast to the bubbling regime, the top surface of the bulged bed does not rupture. A granular jet cannot be observed. Hence, the resulting crater in this regime is symmetric and smooth without a central hump or a surrounding rim (Fig. 3e). At even larger $d$ above 80 mm, due to the high hydrostatic pressure of granular bed, the underground bubbles created by explosion cannot obviously deform the surface. Pressurized air breaks into small bubbles and diffuses through the interstices between particles. Craters do not form at such a deep burial depth.

The four explosion regimes can be also observed at other explosion pressures. Figure 5 shows the phase diagram of different explosion regimes at different $P$. The boundaries of different regimes are approximately linear. With increasing $P$, the transitions between different regimes shift to higher burial depths. The regime of eruption cratering becomes wider at higher $P$ across a larger range of $d$.

Across different regimes, we find that the formation of granular corollas results in pronounced thick rims around explosion craters, whereas the falling of granular jets gives rise to irregular humps at the center of explosion craters. The volume of the humps correlates with the size of the jets. These general observations between the dynamics of granular flows and the surface structures of resulting craters should also apply to granular impact cratering.

### 3.2 Crater diameter

We measure the diameter of explosion craters, $D$, at different explosion regimes. Figure 6 shows $D$ as a function of burial depth $d$ at different explosion pressures $P$ and duration times $\tau$. The crater size increases with $d$ in the air-blow and eruption regime and decreases in the bubbling and sink-in regime. An optimal burial depth near the transition between the eruption and the bubbling regime can be observed, which gives the largest crater diameter at a given $P$ and $\tau$. The optimal $d$ increases with increasing $P$ and $\tau$. Note that the nozzle in our experiments points upwards again gravity. It is likely that $D(d)$ for nozzles pointing to different directions show quantitatively different results. Nevertheless, we expect that the non-monotonic trend of $D(d)$ is robust and independent of the direction of the nozzle.

Interestingly, for small $d$, $D(d)$ collapses into a master curve, showing an approximately linear relation with only weak dependences on $\tau$ and $P$. Such a universal relation indicates a geometric origin of the size of explosion craters at small $d$. We propose a simple model here to explain this interesting observation. Direct imaging from 2D explosion indicates that when the valve is open, an underground bubble containing pressurized air is formed.[21] The bubble first grows radially and isotropically at short times and then elongates faster upward towards the surface of the granular bed, resulting in an oval bubble at later times (Fig. 7). Once the top of the bubble breaks through the surface of the bed, a shortcut between the nozzle and the ambient air establishes. From this breaking moment onward, the pressurized air would go directly from the nozzle to the atmosphere without further pushing on the granular bed sideways. Hence, a longer $\tau$ beyond the breaking moment would not affect the size of craters any further, leading to the weak dependence on $\tau$ at small $d$. Furthermore, the width of the bubble reaches maximum at the breaking moment, which directly determines the diameter of the resulting crater. Note that at larger $d$, the valve is close before the bubble reaches the surface of the bed. Hence, the argument above cannot be applied at large $d$.

Quantitatively, we approximate the speed of the expansion of the bubble using the local flux of the pressurized air.[30,31] Within the pressure range of our experiments, the Reynolds number (Re) of the air flow in the granular bed ranges from 500 up to 2000 at a typical burial depth of 30 mm (Appendix A), which is in the laminar-turbulent transition regime. The velocity of the air flux can be calculated using the Ergun equation,[32]

$$\frac{\mathrm{d}r}{\mathrm{d}t} = -\alpha + \sqrt{\alpha^2 + \beta \frac{\Delta P}{h}}, \tag{1}$$

where

$$\alpha = \frac{300\eta(1-\varepsilon)}{7\rho D_p} \ \text{and} \ \beta = \frac{4D_p\varepsilon^3}{7\rho(1-\varepsilon)}.$$

Here, $D_p = 90$ μm is the mean particle diameter and $\varepsilon = 0.4$ is the void fraction. $\eta = 1.81 \times 10^{-5}$ Pa·s is the viscosity of air at room temperature. $\Delta P = P - P_0$ is the pressure difference between the pressurized air and the ambient atmospheric pressure $P_0$, which is constant for a given explosion event. We estimate the density of air, $\rho$, at the arithmetic mean of the pressures $(P + P_0)/2$. $r$ indicates the radial position of the boundary of the bubble, where the origin is set at the opening of the nozzle with the positive $x$ pointing to the right and the positive $y$ pointing upward against gravity (Fig. 7). $h$ is the depth of the bed at the position $r$. Based on Eq. (1), the growth rate of the bubble is larger for higher $P$ and shallower $h$. At short times, $r$ is small. $h \approx d$. The growth of the bubble is spatially isotropic. As the bubble becomes larger, the growth of the bubble accelerates upwards towards the surface, where $h$ is small. Along the vertical ($+y$) direction, we can approximate $h \approx d - r$ in the simple model, ignoring the deformation of granular surface due to the expansion of the bubble, which is significant only towards the end of the breaking process. An integration of Eq. (1) gives an estimate of the breaking time when the bubble reaches the surface,

$$t = \int_0^d \frac{\mathrm{d}r}{-\alpha + \sqrt{\alpha^2 + \beta\Delta P/(d-r)}}.$$

With $h = d$, the propagation speed of the bubble along the horizontal ($x$) direction is constant (Fig. 7). Thus, the half width of the bubble at the breaking point is

$$w = \left(-\alpha + \sqrt{\alpha^2 + \beta\frac{\Delta P}{d}}\right)t = \left(-\alpha + \sqrt{\alpha^2 + \beta\frac{\Delta P}{d}}\right)\int_0^d \frac{\mathrm{d}r}{-\alpha + \sqrt{\alpha^2 + \beta\Delta P/(d-r)}}. \quad (2)$$

Within the pressure range of our experiments, $(\beta\Delta P/d)^{1/2} \gg \alpha$. We can approximate Eq. (2) as

$$w = \sqrt{\frac{\beta\Delta P}{d}}t = \sqrt{\frac{\beta\Delta P}{d}}\int_0^d \sqrt{\frac{d-r}{\beta\Delta P}}\,\mathrm{d}r = \frac{2}{3}d,$$

which is independent of $P$ and $\tau$. Indeed, numerically solving Eq. (2), we find $w = 0.66d$, showing a very weak dependence on $P$ and independent of $\tau$. The diameter of the crater can then be estimated as $D = 2w$. Hence, $D$ is linearly proportional to the burial depth following $D \approx 1.32d$, qualitatively agreeing with the experiments (Fig. 6). Physically, although a higher explosion pressure leads to a faster propagation of the bubble along the horizontal direction, the bubble also breaks sooner and therefore gives a shorter propagation time. The two competing effects cancel out. As a result, the maximal bubble width and therefore the crater diameter depend only on the geometric factor $d$ and is independent of $P$. We should emphasize that the linear relation between $D$ and $d$ also holds for air flows in the laminar regime at low Re and in the highly turbulent regime at very high Re (Appendix A). The quantitative deviation between the model prediction and the experiments probably arises from the loose packing of granular particles near the bed surface. Particles near the surface are blown away by the strong air flow in open air, which gives non-zero $D$ even for surface explosion with $d = 0$. Such an effect is not considered in our simple model of densely packed bed. In addition, the avalanche of crater walls and the refill of craters by granular corollas and jets towards the end of explosion processes also reconstruct explosion craters and modify $D(d)$.[24,33]

### 3.3 Crater size scaling

As the most important characteristics of granular cratering processes, the scaling of crater size with cratering energy has always been the focus of previous studies.[17,18] For the impact of solid spheres at low energy, the diameter of impact craters follows $D \sim E^{1/4}$, where $D$ is the diameter of craters and $E$ is impact energy.[34,35] For the impact of liquid drops, smaller exponents between 1/6 and 1/4 have been reported.[2,4,36] For high-energy explosion cratering, the size of craters follows the energy scaling $D \sim E^{1/3}$,[37,38] although a modified scaling of $D \sim E^{1/3.4}$ has also been suggested.[7] In general, with a selection of independent variables related to the material properties of explosives and granular media, a simple dimensional analysis shows that the size of explosion craters is bounded by the 1/3 and 1/4 scaling rules.[37] The analysis also shows that if the 1/3 scaling holds, the size of craters should be independent of the gravitational acceleration $g$. At low energy, explosion cratering near surface or in quasi-2D granular bed shows various different scaling rules,[20-23] as discussed in the introduction. It is important to examine how the scaling depends on the dimensionality and the burial depth of explosion events.

We study the energy scaling of explosion craters in 3D experiments with different burial depths at low explosion energy. To calculate the explosion energy $E$, we need to consider the dynamics of high-pressure gas flowing out of a large vessel through a narrow nozzle. Landau and Lifshitz have discussed this problem in their renowned courses of theoretical physics (Sec. 97 in Ref. 39). The velocity of the air flux existing the nozzle is given by

$$v = \sqrt{\frac{\gamma P}{\rho}} \left(\frac{2}{\gamma+1}\right)^{(1+\gamma)/2(\gamma-1)},$$

where $\gamma = 1.4$ is the adiabatic index of air at room temperature $T = 300$ K and $P$ is the pressure in the vessel. Landau and Lifshitz further showed that the pressure drop in the nozzle cannot be greater than $P - P^*$, where $P^*$ is the critical pressure

$$P^* = P \left(\frac{2}{\gamma+1}\right)^{\frac{\gamma}{\gamma-1}} = 0.53P.$$

for air at room temperature. If the external pressure outside the nozzle is smaller than $P^*$, the pressure of the air at the opening of the nozzle is fixed at $P^*$. Since the air pressure in our granular bed before explosion is the atmospheric pressure $P_0$ with $P_0 \ll P^*$, the pressure at the exit of the nozzle is fixed at $P^*$ in our study. We have experimentally verified this prediction in our setup in the previous study.[21]

The energy of a pressurized air pulse is composed of two parts, i.e., the kinetic energy of the air pulse existing the nozzle, $K$, and the internal energy of the air pulse, $U$. We calculate $K$ simply by

$$K = \frac{1}{2}mv^2,$$

where $m$ is the total mass of the air pulse going through the nozzle during explosion,

$$m = \rho v A \tau. \tag{3}$$

Here, $A = 0.25$ mm$^2$ is the cross-section area of the opening of the nozzle. We compute the density of air $\rho = PM/RT$ from the ideal gas law with the molar mass of air $M = 29$ g/mol and the ideal gas constant $R = 8.314$ J/(K·mol).

We calculate the internal energy of the pressurized air pulse, $U$, based on its mechanical $PV$ work.[21] From Eq. (3), the total volume of the pressurized air when leaving the nozzle is

$$V^* = \frac{mRT}{P^*M},$$

where $P^*$ is the critical pressure at the exit of the nozzle. We assume that the fast explosion is an adiabatic process, where the air pulse of volume $V^*$ at pressure $P^*$ expands into volume $V_0$ at the atmospheric pressure $P_0$. $V_0$ is thus given by

$$V_0 = \left(\frac{P^* V^{*\gamma}}{P_0}\right)^{1/\gamma}.$$

The internal energy of the air pulse can then be calculated from the mechanical work of the expansion,

$$U = \int_{V^*}^{V_0} P \mathrm{d}V = \int_{V^*}^{V_0} \frac{P^* V^{*\gamma}}{V^\gamma} \mathrm{d}V = P^* V^{*\gamma} \left(\frac{V_0^{1-\gamma} - V^{*1-\gamma}}{1-\gamma}\right). \tag{4}$$

Within the pressure range of our experiments, we find that $K$ is at most 20% of $U$. Hence, explosion cratering is dominant by the internal energy and, therefore, the expansion of the pressurized air. The kinetic energy plays the leading role only in the shallow air-blow regime, where the expansion of the air pulse is far from complete when reaching the surface of the granular bed. To construct a minimal model with the dominant factor, we approximate the explosion energy $E$ by the internal energy alone,

$$E \approx U. \tag{5}$$

This approximation is validated by the good collapse of energy-scaled data shown in Fig. 8. Having the nozzle point sideways or even downwards can further reduce the influence of kinetic energy on cratering dynamics. The energy of our explosions is between 2.5 J and 18 J from Eq. (5). In comparison, the potential energy of a single particle over a typical burial depth $d = 40$ mm is $3.7 \times 10^{-10}$ J.

Studies at high explosion energies above $5 \times 10^8$ J showed that the diameter of explosion craters follows a 1/3 scaling, $D \sim E^{1/3}$.[7,15] The scaling holds for explosions at similar depths, i.e., at depths scaled according to the same 1/3 scaling rule. Motivated by this previous finding, we plot $D/E^{1/3}$ versus $d/E^{1/3}$ for explosion craters of our low-energy explosions (Fig. 8). Surprisingly, although there are more than seven orders of magnitude difference in energy scales, the 1/3 scaling rule still applies in our experiments and leads to a good collapse of data. The result suggests that both the optimal burial depth and the maximal size of explosion craters scale with the explosion energy as $\sim E^{1/3}$. The collapse of the data justifies the derivation of the explosion energy of pressurized air pulses (Eq. (5)). The scaling rule in 3D granular media is different from that observed in explosion in quasi-2D granular media, where the 1/3 scaling rule applies only in the bubbling regime but fails in the eruption regime.[21] The result is consistent with the $E^{0.3}$ scaling found in the previous study of near-surface explosions at $d = 0$.[20] Note that the self-similar depth $d/E^{1/3}$ is independent of $E$ when $d = 0$. The scaling is in sharp contrast to the 1/4 scaling rule of low-energy impact cratering. Based on a simple energy argument, if the cratering energy $E$ is all converted to the potential energy of granular particles in the crater $\rho_g L^3 g L$, where $L$ is the linear dimension of the crater and $\rho_g$ is the density of granular particles, then $L \sim E^{1/4}$.

Finally, we also attempt to rescale the crater diameter directly using dimensional analysis. The relevant physical quantities in our cratering processes include the explosion pressure $P$, the duration time $\tau$, the burial depth $d$, the density of air $\rho_a$, the cross-section area of the nozzle $A$, the density of granular particles $\rho_g$, the size of granular particles $a$ and the gravitational acceleration $g$. The $\Pi$-theorem dictates that the number of dimensionless groups is equal to the total number of governing parameters minus the number of governing parameters with independent dimensions.[40] Thus, we should have six dimensionless groups, which can be written in a scaling relation as:

$$\frac{D}{d} = f\left(\frac{d}{\tau\sqrt{P/\rho_g}}, \frac{a}{\tau\sqrt{gd}}, \frac{a}{d}, \frac{\rho_a}{\rho_g}, \frac{a^2}{A}\right).$$

The ratio $a^2/A = 0.03$ is a constant in our experiments. Since $\rho_a << \rho_g$, we can ignore the dependence of $\rho_a/\rho_g$. $\rho_g = 2.52$ g/cm$^3$ for soda-lime glass beads. Similarly, $a << d$ for our experiments with $d \geq 5$ mm. Thus, we can also ignore the ratio $a/d$ in the relation too. The second dimensionless group, $I_p \equiv a/\tau(gd)^{1/2}$, can be seen as the inertial number of granular flows, $I_p = \frac{\dot{\gamma}a}{\sqrt{p/\rho_g}}$, where the pressure on granular particles is provided by the hydrostatic pressure $p = \rho_g gd$ and the rate of strain $\dot{\gamma} = 1/\tau$.[41] A fast deformation of granular bed occurs at early times when the valve is open over a time interval of $\tau$ (see Fig. 7). Thus, $1/\tau$ gives the rate of deformation of granular bed, which is equivalent to the average strain-rate on granular particles at the early times of the explosion process. The momentum/energy transferred to the deformed bed in this process determines the behavior of granular flows and eventually the morphology of granular craters. $I_p$ is between 0.002 and 0.02, depending on $d$ and $\tau$ chosen in our experiments. This range of $I_p$ suggests that the granular flows in the cratering progress are in the dense-flow regime.[42] Lastly, the first term, $I_s \equiv d/\tau(P/\rho_g)^{1/2}$, has the same structure as the inertial number. Here, the length scale is provided by the burial depth instead of the particle size and the pressure is given by the explosion pressure instead of the hydrostatic pressure. $I_s$ ranges from 0.003 to 0.1 in our study.

To simplify the scaling relation further, we recall the conclusion from the previous dimensional analysis of explosion cratering by explosives,[37] where it was found that the 1/3 energy scaling shown in Fig. 7 excludes the dependence on $g$. Thus, we first assume that the rescaled diameter is independent of $I_p$. A simple scaling relation is then obtained:

$$\frac{D}{d} = f\left(I_s \equiv \frac{d}{\tau\sqrt{P/\rho_g}}\right). \tag{6}$$

At a fixed $\tau = 20$ ms, Eq. (6) provides an excellent scaling relation to collapse our experimental results (Fig. 9a). However, this simple scaling relation is less satisfying for data at different $\tau$. Including the small yet non-zero inertial number allows a better collapse (Fig. 9b). The final scaling relation follows

$$\frac{D}{d} = f\left(I_s, I_p \equiv \frac{a}{\tau\sqrt{gd}}\right) = f\left(I_s^\alpha I_p^\beta\right), \tag{7}$$

where the power exponents are $\alpha = -1.7$ and $\beta = 1$. A kink in the scaling relation can be observed around $I_s^\alpha I_p^\beta = 25$, which corresponds to the transition between the air-blow and the eruption regime. Thus, the result further supports the classification of different explosion regimes based on high-speed photography and the morphology of explosion craters.

## 4 Conclusion

We studied explosion cratering in 3D granular media. Low-energy explosions were generated in our lab using short pulses of pressurized air with controlled pressures and duration times. We analyzed explosion cratering dynamics and the morphology of the resulting craters at different burial depths. Four different explosion regimes, i.e., the air-blow, eruption, bubbling and sink-in regime, were identified when we increased the burial depth of explosion. A phase diagram of the four regimes was then drawn as a function of explosion pressures and burial depths. A general relation between the dynamics of granular flows and the surface morphologies of resulting craters is discussed. Moreover, we also measured the diameters of explosion craters, which shows a non-monotonic dependence on burial depths. An explosion crater reaches its maximum at an optimal intermediate burial depth, which increases with both explosion pressures and durations. In contrast, crater diameters are independent of explosion pressures and durations at small burial depths. Such an unexpected finding was explained by a simple model of bubble expansion. Finally, we also studied the energy scaling of crater sizes based on the energy formula of pressurized air pulses derived previously. Crater diameters show a cube-root scaling with explosion energy at similar burial depths. The same scaling rule also holds for high-energy explosions despite the more than seven orders of magnitude difference in energy scales. Using dimensional analysis, we also discussed the scaling relation between the rescaled crater diameter and the inertial number of granular flows. As such, our experiments illustrate the rich dynamics of explosion cratering in 3D granular media and reveal the universality of the energy scaling rule of explosion cratering over a wide range of explosion energies. The work paves the way for studying the analogy between explosion cratering and impact cratering at low-energy scales.

## Conflicts of interest

There are no conflicts to declare.

## Acknowledgments

We thank Leonardo Gordillo for helping with experiments. The work is supported by NSF CAREER DMR-1452180.

## Appendix A: Air flows in granular bed

Fluid flows in packed granular bed have been well studied in chemical engineering due to their importance for catalytic reactors and for separation processes.[32] The behavior of air flows in a granular bed depends on the bed Reynolds number defined as

$$\text{Re} = \left(\frac{\rho V_0 D_p}{\eta}\right)\left(\frac{1}{1-\varepsilon}\right),$$

where $D_p$ is the mean particle diameter, $\varepsilon$ is the void fraction, $\eta$ is the air viscosity, $\rho$ is the air density and $V_0$ is the superficial velocity of air flux through the bed.

At low Re < 10, fluid flows are in the laminar regime described by the Blake-Kozeny equation,[32] which is equivalent to Darcy's law:

$$V_0 = -\frac{k}{\eta}\nabla P = \frac{D_p^2}{150}\frac{\varepsilon^3}{(1-\varepsilon)^2}\frac{1}{\eta}\frac{\Delta P}{h}, \tag{8}$$

where $\kappa = \frac{D_p^2}{150}\frac{\varepsilon^3}{(1-\varepsilon)^2}$ is the permeability of the granular bed and $\Delta P/h$ gives the local pressure gradient. At high Re > 1000, air flows are in the highly turbulent regime described by the Burke-Plummer equation,

$$V_0 = \left(\frac{4}{7}\frac{D_p}{\rho}\frac{\varepsilon^3}{1-\varepsilon}\frac{\Delta P}{h}\right)^{1/2}. \tag{9}$$

In the laminar-turbulent transition regime between the two limits, air flows can be described by the Ergun equation with

$$V_0 = -\alpha + \sqrt{\alpha^2 + \beta\frac{\Delta P}{h}},$$

where $\alpha$ and $\beta$ are given in the main text (see Eq. (1)). The Ergun equation reduces to the Blake-Kozeny equation at small $V_0$, whereas it transforms to the Burke-Plummer equation at large $V_0$. Hence, the equation provides a quantitative description of fluid flows in granular bed over a broad range of Re beyond the transition regime (see Fig. 6.4-2 of Ref. 32).

We estimate Re by calculating $V_0$ in each regime and checking the consistency of Re. With a typical burial depth $h = d = 30$ mm and the pressure range between $2.1 \times 10^6$ Pa and $8.3 \times 10^6$ Pa, $V_0$ is on the order of 5 m/s, which gives Re in the range of 500 to 2000. Thus, the air flows in our experiments are in the transition regime described by the Ergun equation. Note that we estimate the density of air at the arithmetic mean of the pressures $(P + P_0)/2$, where $P$ is the pressure in the reservoir and $P_0$ is the atmospheric pressure. The pressure difference is $\Delta P = P - P_0$. To be more accurate, we should use the pressure at the exit of the narrow nozzle, which is fixed at $0.53P$ based on the Landau's theory discussed in the main text.[39] Nevertheless, since the half-width of the bubble is independent of pressure in the pressure range of our experiments, the conclusion of our model does not change when we use $0.53P$ instead of $P$.

The linear relation between $D$ and $d$ is obtained in both the laminar regime and the highly turbulent regime. At low Re in the laminar regime, $dr/dt$ (= $V_0$) is proportional to the pressure gradient $\Delta P/h$ from Darcy's law (Eq. (8)). The same analysis used in the main text leads to $w = d/2$ and $D = d$, independent of $\tau$ and $P$. At very high Re in the highly turbulent regime, $dr/dt$ is proportional to $(\Delta P/h)^{1/2}$ from the Burke-Plummer equation (Eq. (9)). The analysis gives a linear relation $w = 2d/3$ and $D = 4d/3$, independent of $\tau$ and $P$, as shown in the main text.

# References:

1    S. M. Som, D. C. Catling, J. P. Harnmeijer, P. M. Polivka and R. Buick, *Nature*, 2012, **484**, 359–362.

2    R. Zhao, Q. Zhang, H. Tjugito and X. Cheng, *Proc. Natl. Acad. Sci. USA*, 2015, **112**, 342–347.

3    S. C. Zhao, R. de Jong and D. van der Meer, *Soft Matter*, 2015, **11**, 6562–6568.

4    Q. Zhang, M. Gao, R. Zhao and X. Cheng, *Phys. Rev. E*, 2015, **92**, 042205.

5    J. P. Johnson and K. X. Whipple, *Earth Surf. Proc. Land.*, 2007, **32**, 1048–1062.

6    A. E. Boudreau, *Aust. J. Earth Sci.*, 1992, **39**, 277–287.

7    H. J. Melosh, *Impact Cratering: A Geologic Process*, Oxford University Press, Oxford, UK, 1989.

8    W. K. Hartmann, M. Malin, A. McEwen, M. Carr, L. Soderblom, P. Thomas, E. Danielson, P. James and J. Veverka, *Nature*, 1999, **397**, 586–589.

9    L. Pawłowski, *The Science and Engineering of Thermal Spray Coatings*, 2nd edn, Wiley, West Sussex, UK, 2008.

10   P. W. Cooper, *Explosives Engineering*, Wiley-VCH, New York, USA, 1996.

11   K. S. Marhadi and V. K. Kinra, *J. Sound Vib.*, 2005, **283**, 433–448.

12   R. Hooke, *Micrographia: or Some Physiological Descriptions of Minute Bodies Made by Magnifying Glasses. With Observations and Inquiries Thereupon*, J. Martyn and J. Allestry, London, UK, 1665.

13   V. R. Oberbeck, *J. Geophys. Res.*, 1971, **76**, 5732–5749.

14   K. A. Holsapple, *Proc. Lunar Planet. Sci. Conf.*, 1980, **11**, 2379–2401.

15   M. D. Nordyke, *J. Geophys. Res.*, 1962, **67**, 1965–1973.

16   M. D. Nordyke, *Ann. Nucl. Energy*, 1975, **2**, 657–660.

17   J. C. Ruiz-Suarez, *Rep. Prog. Phys.*, 2013, **76**, 066601.

18   D. van der Meer, *Annu. Rev. Fluid Mech.*, 2017, **49**, 463–484.

19   H. Katsuragi, *Physics of Soft Impact and Cratering*, Springer Japan, Tokyo, 2016.

20   F. Pacheco-Vazquez, A. Tacuma and J. O. Marston, *Phys. Rev. E*, 2017, **96**, 032904.

21   M. Gao, X. Liu, L. P. Vanin, T.-P. Sun, X. Cheng and L. Gordillo, *AIChE J.*, 2018, **64**, 2972–2981.

22     F. E. Loranca-Ramos, J. L. Carrillo-Estrada and F. Pacheco-Vazquez, *Phys. Rev. Lett.*, 2015, **115**, 028001.

23     J. O. Marston and F. Pacheco-Vazquez, *Phys. Rev. E*, 2019, **99**, 030901.

24     J. F. Boudet, Y. Amarouchene and H. Kellay, *Phys. Rev. Lett.*, 2006, **96**, 158001.

25     G. Caballero, R. Bergmann, D. van der Meer, A. Prosperetti and D. Lohse, *Phys. Rev. Lett.*, 2007, **99**, 018001.

26     X. Cheng, G. Varas, D. Citron, H. Jaeger and S. R. Nagel, *Phys. Rev. Lett.*, 2007, **99**, 188001.

27     J. R. Royer, E. I. Corwin, B. Conyers, A. Flior, M. L. Rivers, P. J. Eng and H. M. Jaeger, *Phys. Rev. E*, 2008, **78**, 011305.

28     J. R. Royer, E. I. Corwin, A. Flior, M.-L. Cordero, M. L. Rivers, P. J. Eng and H. M. Jaeger, *Nat. Phys.*, 2005, **1**, 164–167.

29     X. Cheng, L. Gordillo, W. W. Zhang, H. M. Jaeger and S. R. Nagel, *Phys. Rev. E*, 2014, **89**, 042201.

30     X. Cheng, R. Smith, H. M. Jaeger and S. R. Nagel, *Phys. Fluids*, 2008, **20**, 123305.

31     J. F. Davidson and D. Harrison, *Fluidised Particles*, Cambridge University Press, Cambridge, UK, 1963.

32     R. B. Bird, W. E. Stewart and E. N. Lightfoot, *Transport Phenomena, 2nd edn.*, Wiley, New York, 2007.

33     E. Lajeunesse, *Phys. Fluids*, 2004, **16**, 2371–2381.

34     J. S. Uehara, M. A. Ambroso, R. P. Ojha and D. J. Durian, *Phys. Rev. Lett.*, 2003, **90**, 194301.

35     A. M. Walsh, K. E. Holloway, P. Habdas and J. R. de Bruyn, *Phys. Rev. Lett.*, 2003, **91**, 104301.

36     H. Katsuragi, *Phys. Rev. Lett.*, 2010, **104**, 218001.

37     K. A. Holsapple and R. M. Schmidt, *J. Geophys. Res.*, 1980, **85**, 7247–7256.

38     H. Sato and H. Taniguchi, *Geophys. Res. Lett.*, 1997, **24**, 205–208.

39     L. D. Landau and E. M. Lifshitz, *Fluid Mechanics, 2nd edn*, Pergamon Press, Oxford, UK, 1987.

40     G. I. Barenblatt, *Scaling*, Cambridge University Press, Cambridge, UK, 2003.

41     P. Jop, Y. Forterre and O. Pouliquen, *Nature*, 2006, **441**, 727–730.

42     G. D. R. MiDi, *Eur. Phys. J. E*, 2004, **14**, 341–365.

**Figures:**

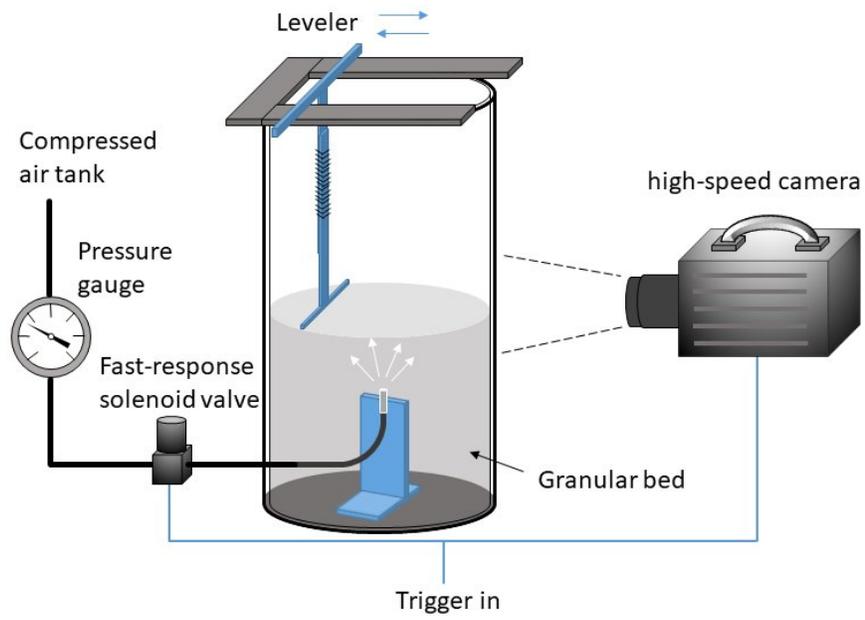

Fig. 1 Schematic of the experimental setup. The schematic is not in scale. The dimensions are stated in the text.

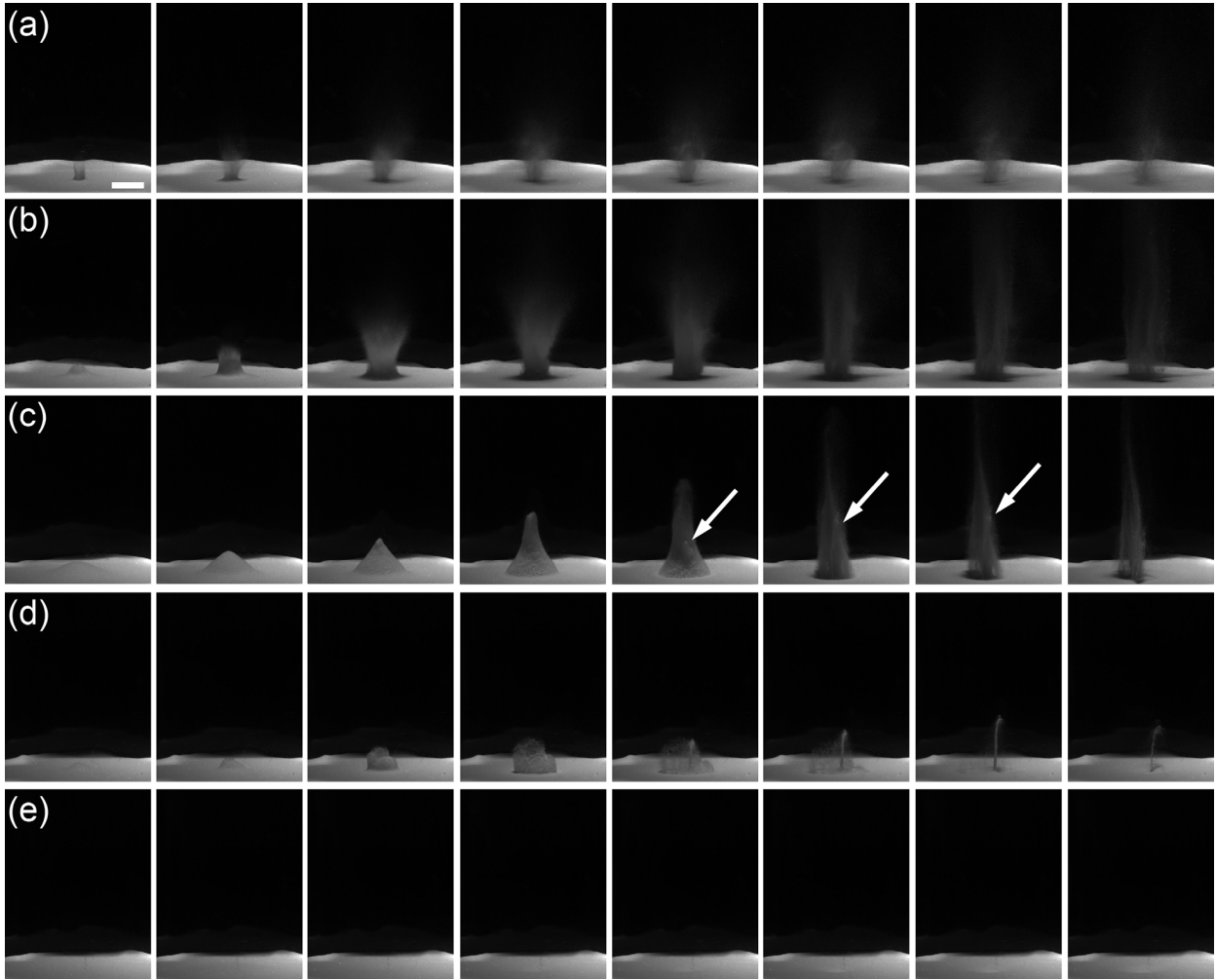

Fig. 2 Explosion dynamics. Snapshots from high-speed videos showing explosion cratering in 3D granular media at different burial depths. Explosion pressure $P = 4.1 \times 10^6$ Pa. Explosion duration $\tau = 20$ ms. Burial depth $d = 5$ mm (a), 20 mm (b), 45 mm (c), 65 mm (d) and 70 mm (e). (a) The air-blow regime. From left to right, the time is $t = 17$ ms, 22 ms, 36 ms, 46 ms, 54 ms, 68 ms, 96 ms and 143 ms. $t = 0$ is the time when the valve is open. (b) The eruption regime. $t = 7$ ms, 10 ms, 23 ms, 41 ms, 54 ms, 114 ms, 138 ms and 172 ms. (c) Transition from eruption to bubbling. $t = 53$ ms, 69 ms, 78 ms, 92 ms, 107 ms, 146 ms, 176 ms and 227 ms. The white arrows indicate a rising granular jet. (d) The bubbling regime. $t = 231$ ms, 237 ms, 250 ms, 272 ms, 297 ms, 308 ms, 332 ms and 402 ms. (e) The sink-in regime. $t = 23$ ms, 79 ms, 200 ms, 361 ms, 374 ms, 419 ms, 488 ms and 588 ms. A shallow crater can be observed on the surface of the bed starting from the fourth frame. All the images have the same scale. The scale bar in (a) is 2 cm.

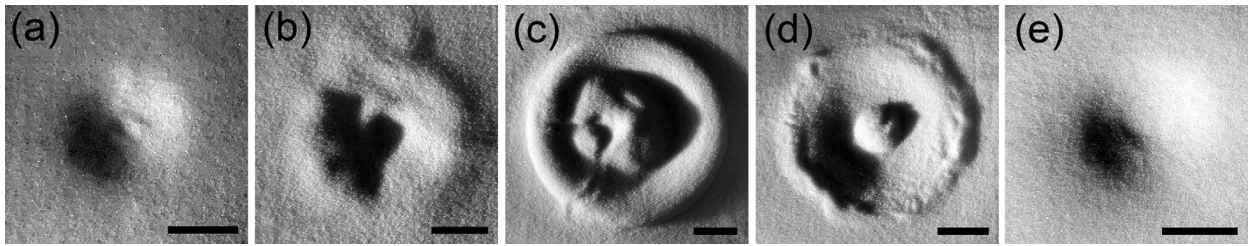

Fig. 3 Explosion craters with increasing burial depths at explosion pressure $P = 4.1 \times 10^6$ Pa and explosion duration $\tau = 20$ ms. Burial depth $d = 5$ mm (a), 20 mm (b), 45 mm (c), 65 mm (d) and 70 mm (e). (a) The air-blow regime. (b) The eruption regime. (c) Transition from eruption to bubbling. (d) The bubbling regime. (e) The sink-in regime. Scale bars = 1 cm.

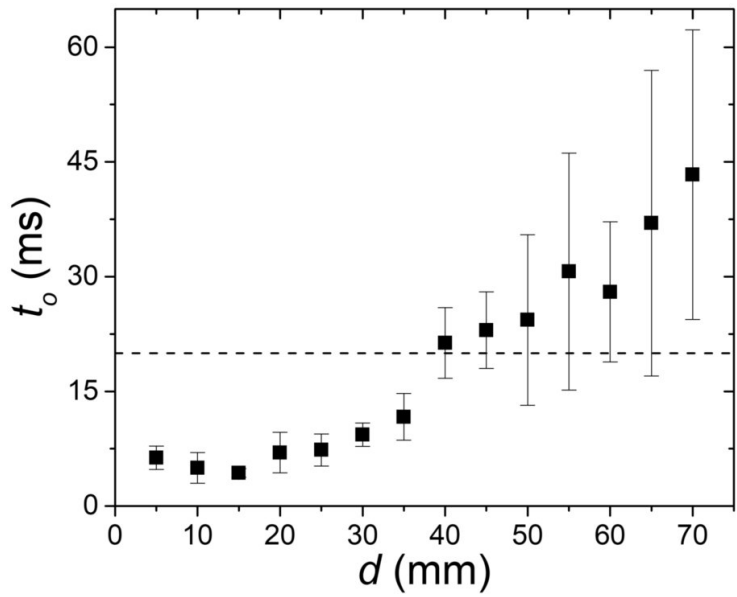

Fig. 4 Onset time $t_o$ versus burial depth $d$. $t_o$ is defined as the time when the surface of granular bed first deforms. $t = 0$ is the time when the valve is open. Explosion pressure $P = 4.1 \times 10^6$ Pa. Explosion duration $\tau = 20$ ms, which is indicated by the horizontal dashed line.

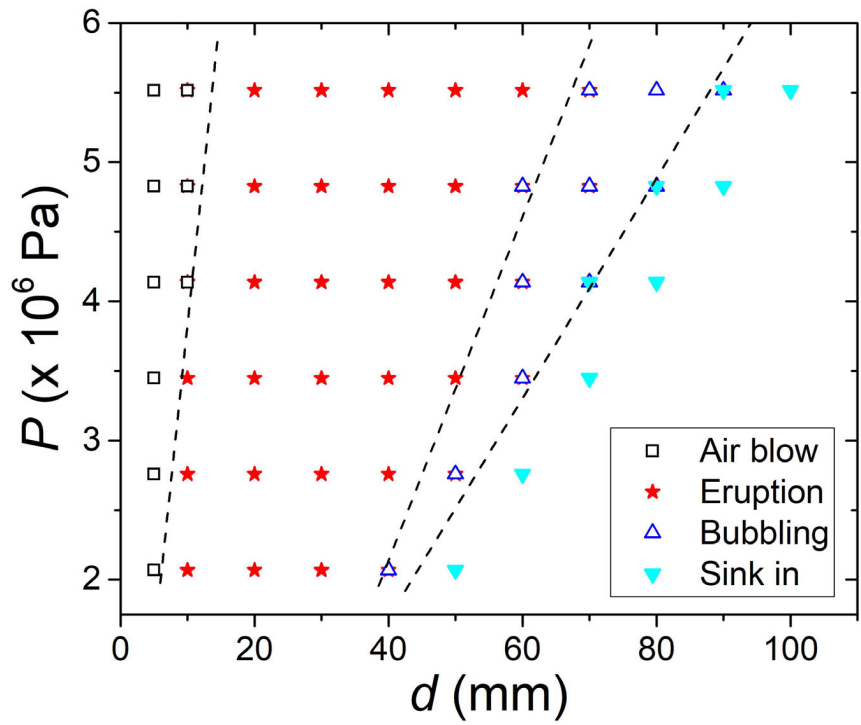

Fig. 5 Phase diagram of explosion cratering at different explosion pressures *P* and different burial depths *d*. Four different regimes are marked in the legend. Dashed lines indicate the boundary between different regimes.

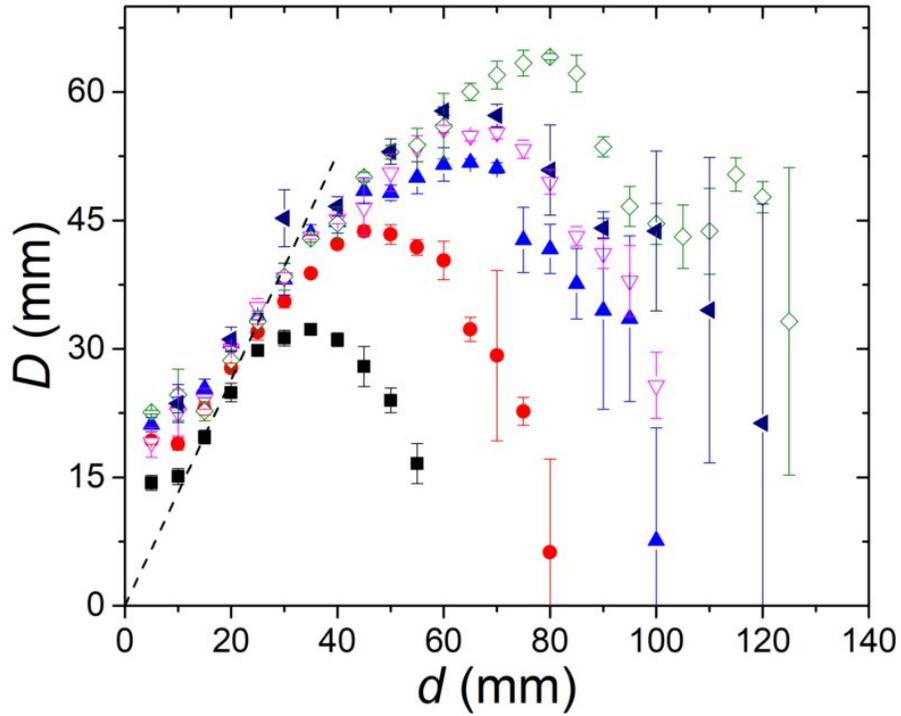

Fig. 6 Diameter of explosion craters, $D$, versus the burial depth, $d$. From the low to high in terms of the maximal crater diameters, the explosion pressures and durations are $P = 2.09 \times 10^6$ Pa and $\tau = 20$ ms (black squares); $P = 4.14 \times 10^6$ Pa and $\tau = 20$ ms (red disks); $P = 6.21 \times 10^6$ Pa and $\tau = 20$ ms (blue up-pointing triangles); $P = 4.14 \times 10^6$ Pa and $\tau = 40$ ms (magenta down-pointing triangles); $P = 8.27 \times 10^6$ Pa and $\tau = 20$ ms (navy left-pointing triangles); and $P = 4.14 \times 10^6$ Pa and $\tau = 60$ ms (olive diamonds). The dashed line indicates $D = 1.32d$ from the model prediction.

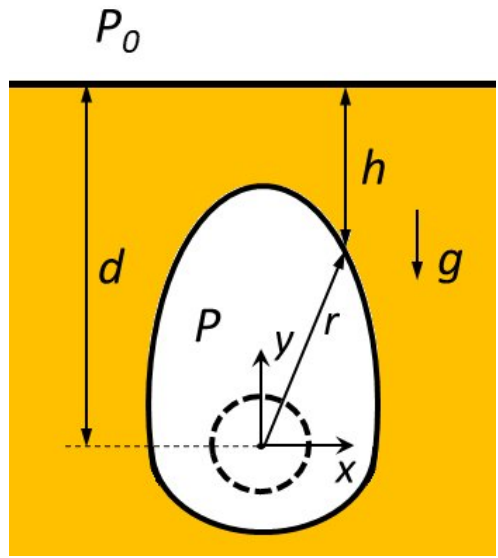

Fig. 7 Schematic showing the expansion of an underground bubble. The dashed circle indicates the shape of the bubble at early times, while the solid oval indicates the shape of the bubble at later times. The expansion is axially symmetric with respect to the $y$ axis. The origin is at the explosion site. The position of the bubble wall $r$, the height $h$, the burial depth $d$, the gravitational acceleration $g$, the pressure of air $P$ and the atmospheric pressure $P_0$ are indicated in the schematic.

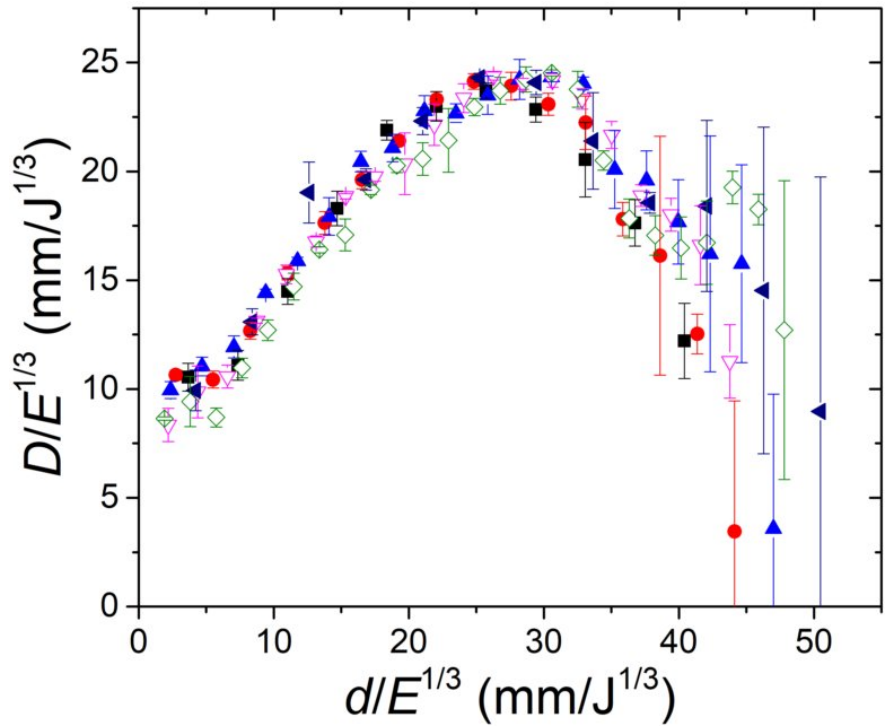

Fig. 8 Energy scaling of the diameter of explosion craters. The rescaled diameter $D/E^{1/3}$ versus the rescaled burial depth $d/E^{1/3}$ for different explosion pressures and durations. $D$ is the crater diameter and $d$ is the burial depth. $E$ is the explosion energy. The symbols are the same as those used in Fig. 6: $P = 2.09 \times 10^6$ Pa and $\tau = 20$ ms (black squares); $P = 4.14 \times 10^6$ Pa and $\tau = 20$ ms (red disks); $P = 6.21 \times 10^6$ Pa and $\tau = 20$ ms (blue up-pointing triangles); $P = 4.14 \times 10^6$ Pa and $\tau = 40$ ms (magenta down-pointing triangles); $P = 8.27 \times 10^6$ Pa and $\tau = 20$ ms (navy left-pointing triangles); and $P = 4.14 \times 10^6$ Pa and $\tau = 60$ ms (olive diamonds).

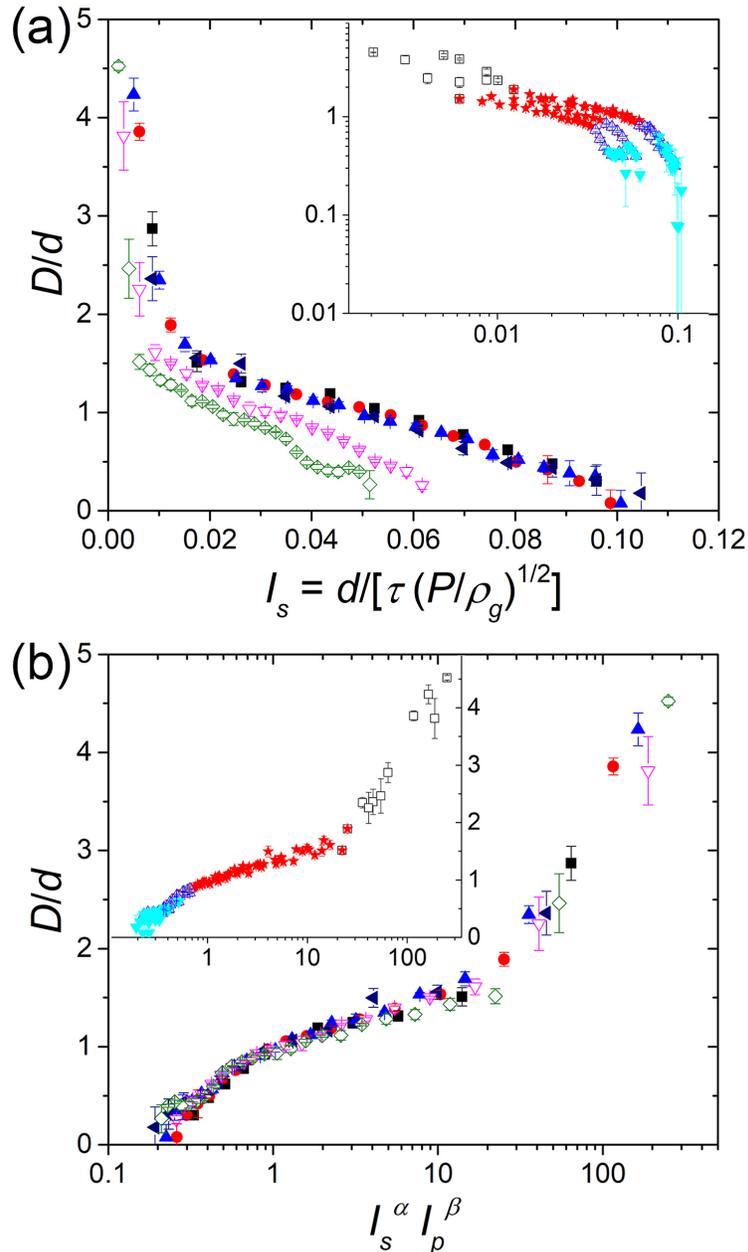

Fig. 9 Scaling relations of crater diameters. (a) Dimensionless crater diameter, $D/d$, versus the dimensionless number, $I_s$. The symbols are the same as those used in Fig. 6: $P = 2.09 \times 10^6$ Pa and $\tau = 20$ ms (black squares); $P = 4.14 \times 10^6$ Pa and $\tau = 20$ ms (red disks); $P = 6.21 \times 10^6$ Pa and $\tau = 20$ ms (blue up-pointing triangles); $P = 4.14 \times 10^6$ Pa and $\tau = 40$ ms (empty magenta down-pointing triangles); $P = 8.27 \times 10^6$ Pa and $\tau = 20$ ms (navy left-pointing triangles); and $P = 4.14 \times 10^6$ Pa and $\tau = 60$ ms (empty olive diamonds). Note that solid symbols are for a fixed explosion duration $\tau = 20$ ms at different explosion pressures and empty symbols are for a fixed explosion pressure $P = 4.1 \times 10^6$ Pa at different explosion durations. Inset shows the same data in terms of explosion regimes. The symbols are the same as those used in Fig. 5: air blow (black squares), eruption (red stars), bubbling (blue up-pointing triangles) and sink-in (cyan down-pointing triangles). (b) $D/d$ versus $I_s$ and the inertial number, $I_p$, with $\alpha = -1.7$ and $\beta = 1$. Inset shows the same data in terms of explosion regimes.